\documentclass[3p,times,twocolumn]{elsarticle}

\usepackage{ecrc}
\usepackage{epsfig}
\usepackage{epstopdf}
\def\bea#1\eea{\begin{align}#1\end{align}} 

\newcommand{\bef}{\begin{figure}[htb]\centering}
\newcommand{\eef}{\end{figure}}

\volume{00}

\firstpage{1}

\journalname{Nuclear Physics B Proceedings Supplement}

\runauth{H. Xing et al.}

\jid{nppp}

\jnltitlelogo{Nuclear Physics B Proceedings Supplement}

\usepackage{amssymb}
\usepackage[figuresright]{rotating}

\begin{document}

\begin{frontmatter}

\dochead{}

\title{Quenching of inclusive and tagged b-jets at the LHC}

\author[HX]{Hongxi Xing}

\author[JH]{Jinrui Huang}

\author[ZK]{Zhong-Bo Kang}

\author[IV]{Ivan Vitev}

\address[HX,JH,ZK,IV]{Theoretical Division, Los Alamos National Laboratory, Los Alamos, NM 87545, USA}

\begin{abstract}
We present theoretical predictions for the nuclear-induced attenuation of the differential cross sections for inclusive and tagged b-jet production in heavy ion collisions at the LHC. We find that for inclusive b-jet production at high transverse momentum the mass effects are negligible, and that the attenuation is comparable to the one observed for light jets. On the other hand, for isolated-photon and B-meson-tagged b-jets the sample of events with heavy quarks produced at the early stages of the collision is greatly enhanced. Thus, these  tagged b-jet final-states have a much more direct connection to the physics of b-quark energy loss. We present theoretical predictions for the quenching of such tagged b-jet events at the LHC and the QGP-induced modification of the related momentum imbalance and asymmetry. We demonstrate that these tagged processes can be used to accurately study the physics of heavy quark production and propagation in dense QCD matter. 
\end{abstract}


\end{frontmatter}

\section{Introduction}
\label{}
Theoretical and experimental advances in understanding the nuclear modification of light hadron and jet production in nucleus-nucleus reactions have been a highlight of the heavy ion programs at RHIC and the LHC. The flavor origin of the final state light hadrons and jets at RHIC and LHC energies is dominated by light quarks and gluons. Therefore, the attenuation of light hadron and jet production in nucleus-nucleus  collisions provides us with the opportunity to investigate the energy loss mechanisms of light flavor partons, arising form their interactions in QCD matter, and to understand the in-medium parton shower formation.

At present, the heavy quark jet and heavy meson production data at RHIC and LHC are still not fully understood. It was presented at this conference that the b-jet attenuation measured by CMS behaves similarly to that of light jets~\cite{CMS}, and the attenuation of open heavy meson measured by STAR is similar to that of light hadrons~\cite{STAR}. From a theoretical point of view, it is well known that  heavy quark radiative energy loss is significantly smaller than that of light quarks due to the dead cone effect in the small-angle emission region \cite{Dokshitzer:2001zm}.  It should lead to smaller suppression of b-jets in comparison to light jets in the small and intermediate $p_T$ region if b-jets originate form prompt b-quarks.  However, this naive expectation  contradicts recent experimental data. Various theoretical models have been proposed to address heavy flavor suppression. In this talk, we discuss the relation between the attenuation of inclusive and tagged b-jets production and the physics of heavy quark production and propagation in dense QCD matter. We demonstrate the significance of understanding the detailed flavor origin of the final-state inclusive b-jets in interpreting the CMS data \cite{Huang:2013vaa}, and propose  to measure the photon-tagged and B-meson-tagged b-jets production in heavy ion collisions \cite{Huang:2015mva}. The tagged processes significantly enhance the fraction of b-jets that originate 
from the prompt b-quarks and lead to much more direct connection to b-quark energy loss. 

\section{Quenching of inclusive and tagged b-jets}
We use Pythia 8 \cite{Sjostrand:2007gs} to evaluate the inclusive and tagged b-jets differential cross sections in p+p collisions at LHC energies. The b-jets are defined using the anti-$k_T$ algorithm with at least one b-quark (or $\rm \bar b$-quark) inside the jet cone. The SlowJet program is used for the jet clustering~\cite{Cacciari:2008gp}. The Pythia simulation has been checked against the data on inclusive and photon-tagged b-jets production at the LHC and the Tevatron, respectively, to validate the p+p baseline.

In order to study the energy loss effects and the medium-induced parton shower, we need detailed information on the flavor origin of b-jets in p+p collisions. The various mechanisms that generate the heavy quark can be identified through the $2 \to 2$ hard partonic scattering in Pythia simulations.  
\bef
\psfig{file=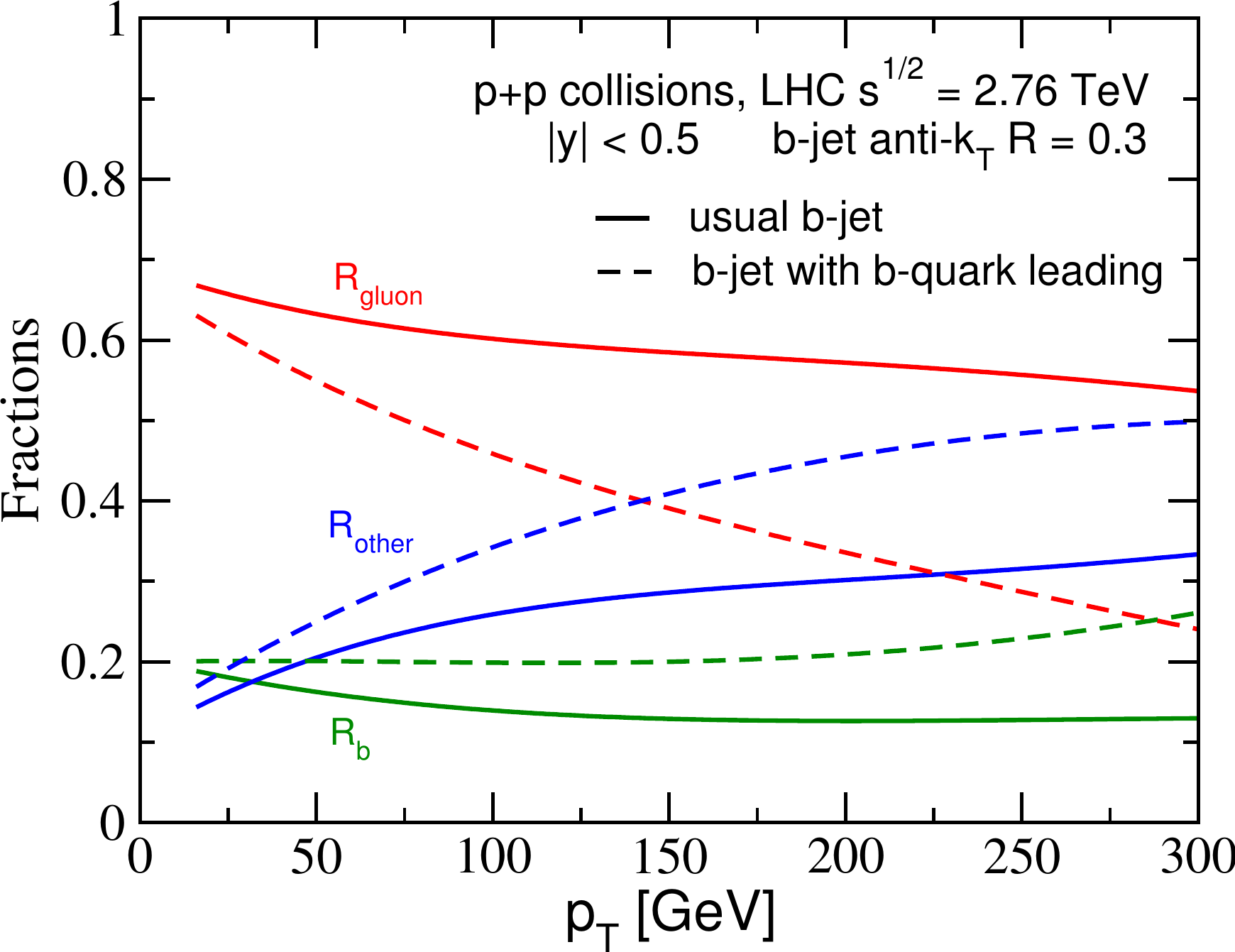, width=0.85\columnwidth}
\caption{The fractional contribution of different subprocesses to  inclusive b-jet production cross section in p+p collisions with $\sqrt{s}=2.76$ TeV.}
\label{b-fraction}
\eef
In Fig. \ref{b-fraction}  we show the fractional contribution of different subprocesses to the commonly defined inclusive b-jet (solid curves) cross section in $\sqrt{s}=2.76$ TeV p+p collisions. We find that the contribution from ``gluon splitting" ($R_g$) is dominant in a wide $p_T$ range, while the prompt b-quark contribution ($R_b$), which has a direct connection to the heavy quark energy loss, is only around $10 -20\% $. This is very different from the earlier expectation that the nuclear modification of inclusive b-jets is leading by the b-quark energy loss in dense QCD matter. Even with the requirement that the contained b-quark is the leading particle inside the jet cone, the dominant contribution still comes from ``gluon splitting" in small $p_T$ region, and light flavor generation ($R_{\rm other}$) in large $p_T$. The fraction from b-quark is always subdominant.

The medium modified b-jet cross section in nucleus-nucleus collisions is calculated as follows:
\begin{equation}
d\sigma^{\rm AA}(R) =\sum_s d\sigma_{(s)}^{pp}(R,\epsilon)\otimes P_{(s)}(\epsilon)|J_{(s)}(R,\epsilon)|\, ,
\end{equation}
where the differential phase space factors are omitted for brevity. $P_{(s)}(\epsilon)$ is the probability distribution that the hard parent parton in color state ${(s)}$ will redistribute a fraction of its energy $\epsilon$ into a medium-induced parton shower \cite{Vitev:2005he}. The phase space Jacobian $|J_{(s)}(\epsilon)|$ accounts for what fraction of the medium-induced parton shower energy is retained inside the measured jet cone, as opposed to``lost" outside \cite{Vitev:2008rz}.
We have included the radiative energy loss as well as the dissipation of energy of the parton shower through collisional interactions \cite{Neufeld:2011yh}. In the simulation, one can estimate that the heavy quark generation time from gluon splitting or light quark parton shower is much smaller than the typical size of the medium~\cite{Huang:2013vaa}. Thus the medium modification of these b-jets should resemble that of a massive gluon or massive quark transversing the QGP. In the following numerical evaluations, the range of masses for the collimated propagating parent parton system is assumed to be  between $m_b$ and $2m_b$ for the purpose of calculating radiative energy loss.

\bef
\psfig{file=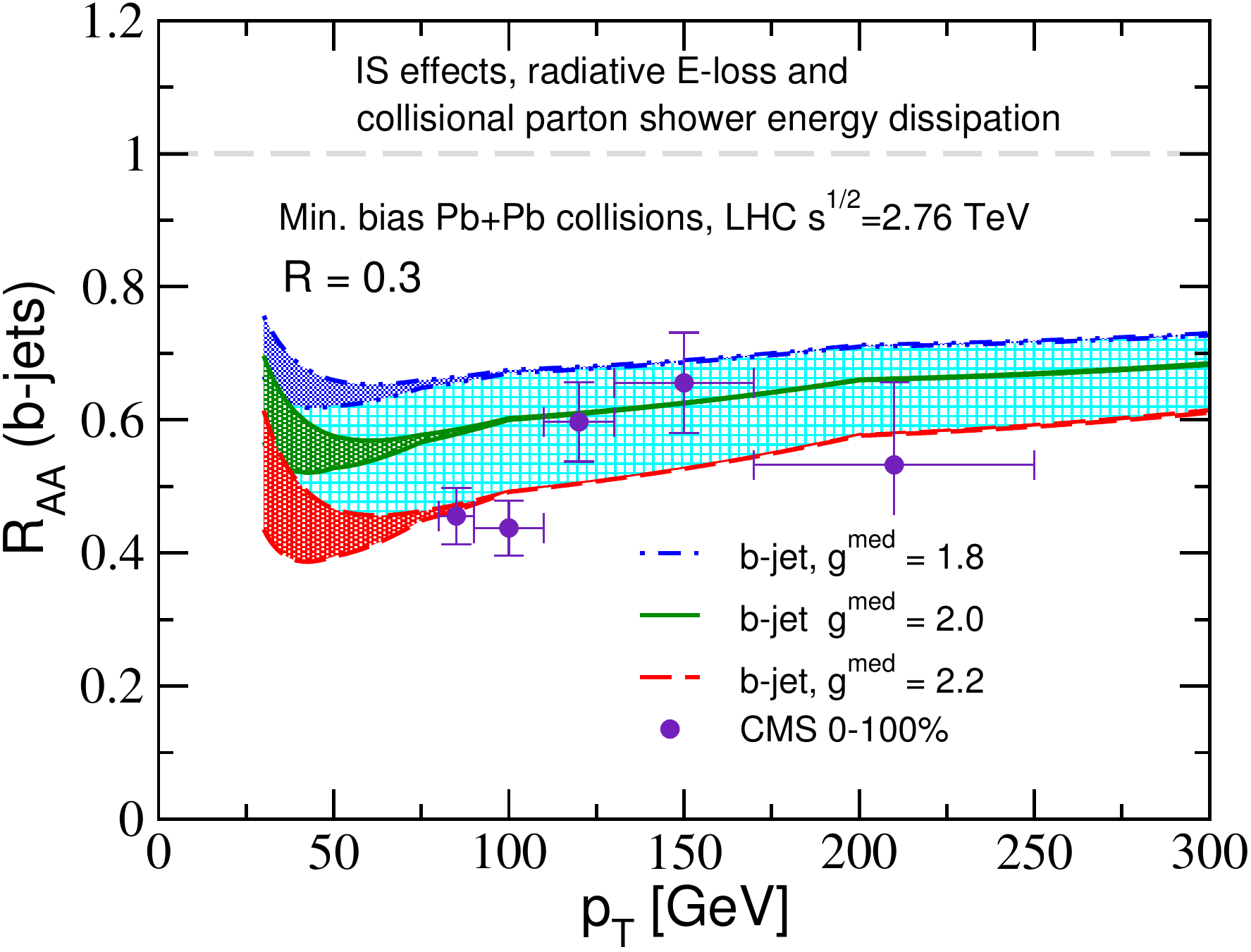, width=0.9\columnwidth}
\caption{The $p_T$-dependent suppression of the inclusive b-jet production cross section in minimum-bias Pb+Pb collisions at the LHC $\sqrt{s}=2.76$ TeV. Data is included form the subsequent CMS collaboration publication \cite{Chatrchyan:2013exa}. }
\label{Raa-b}
\eef

The difference of the energy distribution of a hard scattering parton in the final state between a vacuum and a medium will lead to a modification of the measured b-jet cross section, which can be quantified through the nuclear modification factor $R_{AA}$. 
Shown in Fig. \ref{Raa-b} is the comparison between our theoretical evaluation and the CMS measurement in minimum bias Pb+Pb collisions at $\sqrt{s} = 2.76$ TeV. We observe that there is agreement between our simulations and the CMS measurements within the uncertainty of the jet-medium coupling strength $g^{\rm med} = 1.8 - 2.2$. It is not surprising that the suppression of b-jets is similar to that of light quark jets, due to the dominant gluon splitting contribution as shown in Fig. \ref{b-fraction} and the disappearance of heavy quark mass effect in the large transverse momentum region, as expected in the coherent energy loss picture (the final-state Landau-Pomeranchuk-Migdal effect).

To enhance the fraction of b-jets that originate from prompt b-quarks and provide constraints on the energy of the parton shower, we extend the calculation of inclusive b-jets to tagged b-jet production in heavy ion collisions.  These channels constrain much better the parton flavor origin of the parton shower that recoils against the tagging particle. In particular, we focus on isolated-photon-tagged and B-meson-tagged b-jets production at $\sqrt{s}=5.1$ TeV, which should be experimentally accessible at the LHC.
\bef
\psfig{file=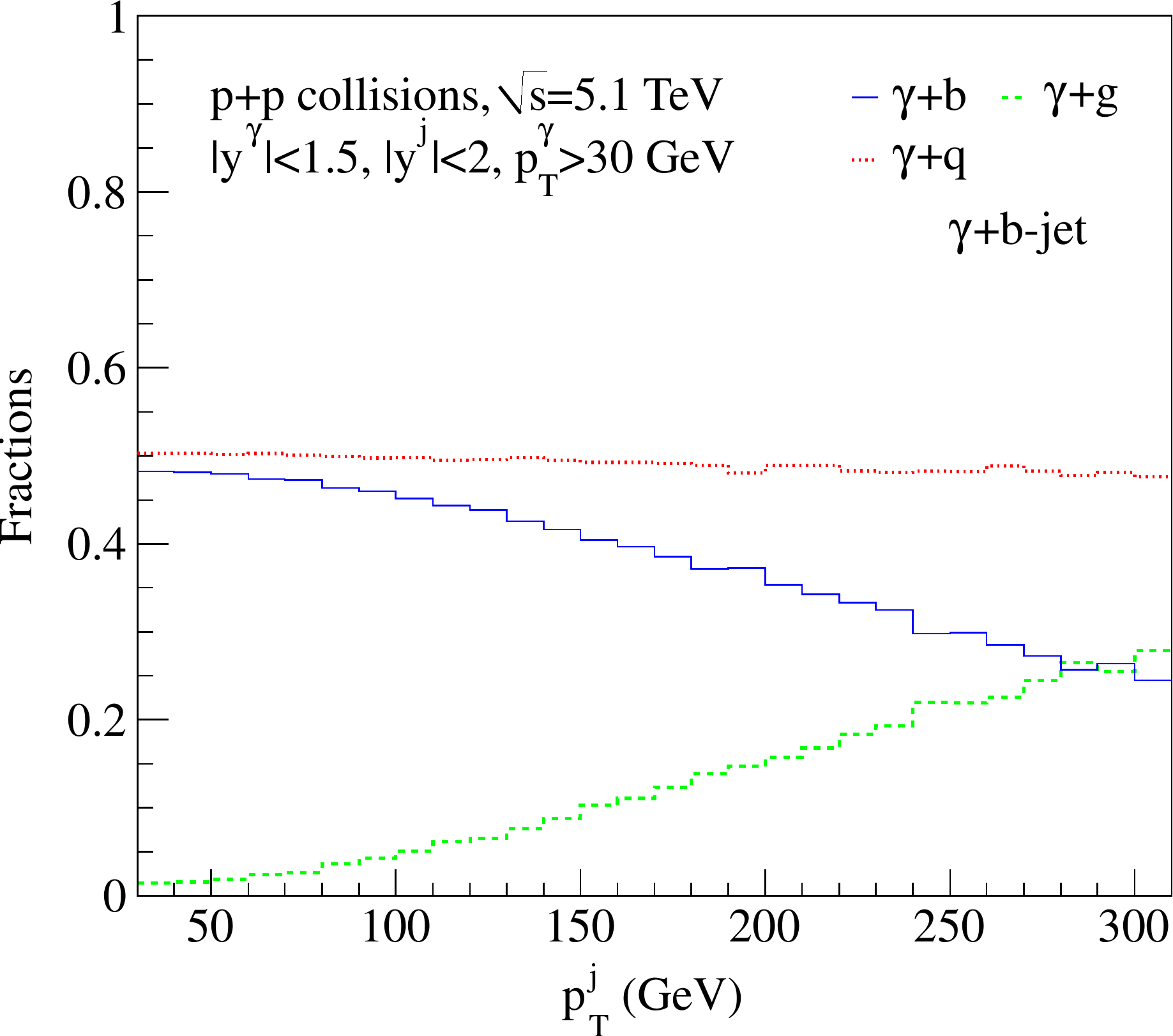, width=0.78\columnwidth}
\hskip 0.1 in 
\psfig{file=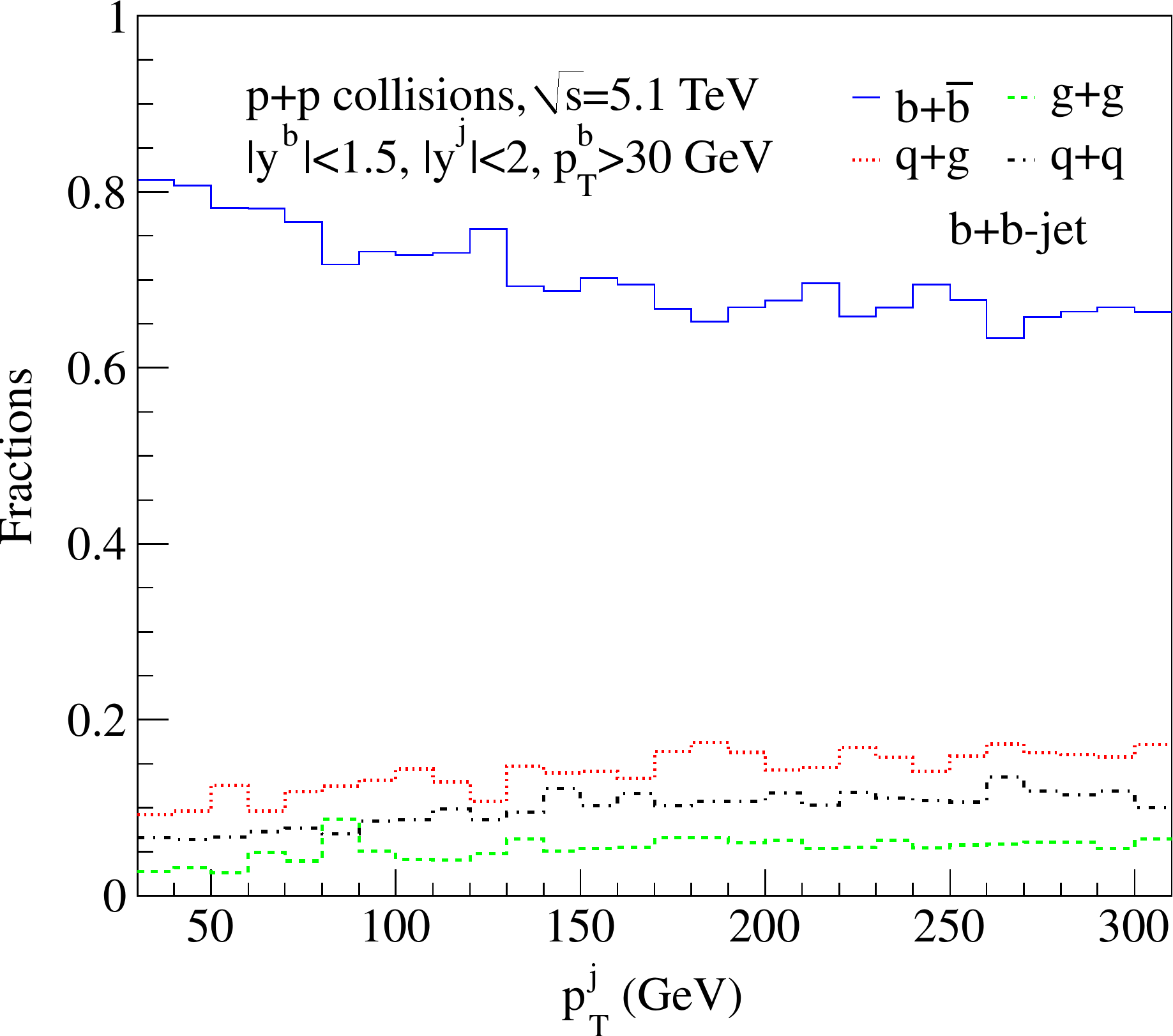, width=0.78\columnwidth}
\caption{The fractional contribution of different subprocess to the photon-tagged (top) and B-meson-tagged (bottom) b-jet production cross sections in p+p collisions with $\sqrt{s}=5.1$ TeV. }
\label{tagged-b-fraction}
\eef
The details of the flavor origin of photon-tagged and B-meson-tagged b-jet are shown in Fig. \ref{tagged-b-fraction}. In photon-tagged b-jet production, the fractional contribution from prompt b-quarks reaches  about $50\%$ when the b-jet transverse momentum are not very large $p_T^j<100$ GeV. This implies that this process has much closer connections to the physical heavy quark energy loss in comparison to inclusive b-jets production. One can further increase the fractional contribution from prompt b-quarks by considering B-meson-tagged b-jets production, where the $b+\bar b$ contribution dominates in the whole kinematic region and provides a tool to directly probe the physical b-quark energy loss mechanism.

\bef
\psfig{file=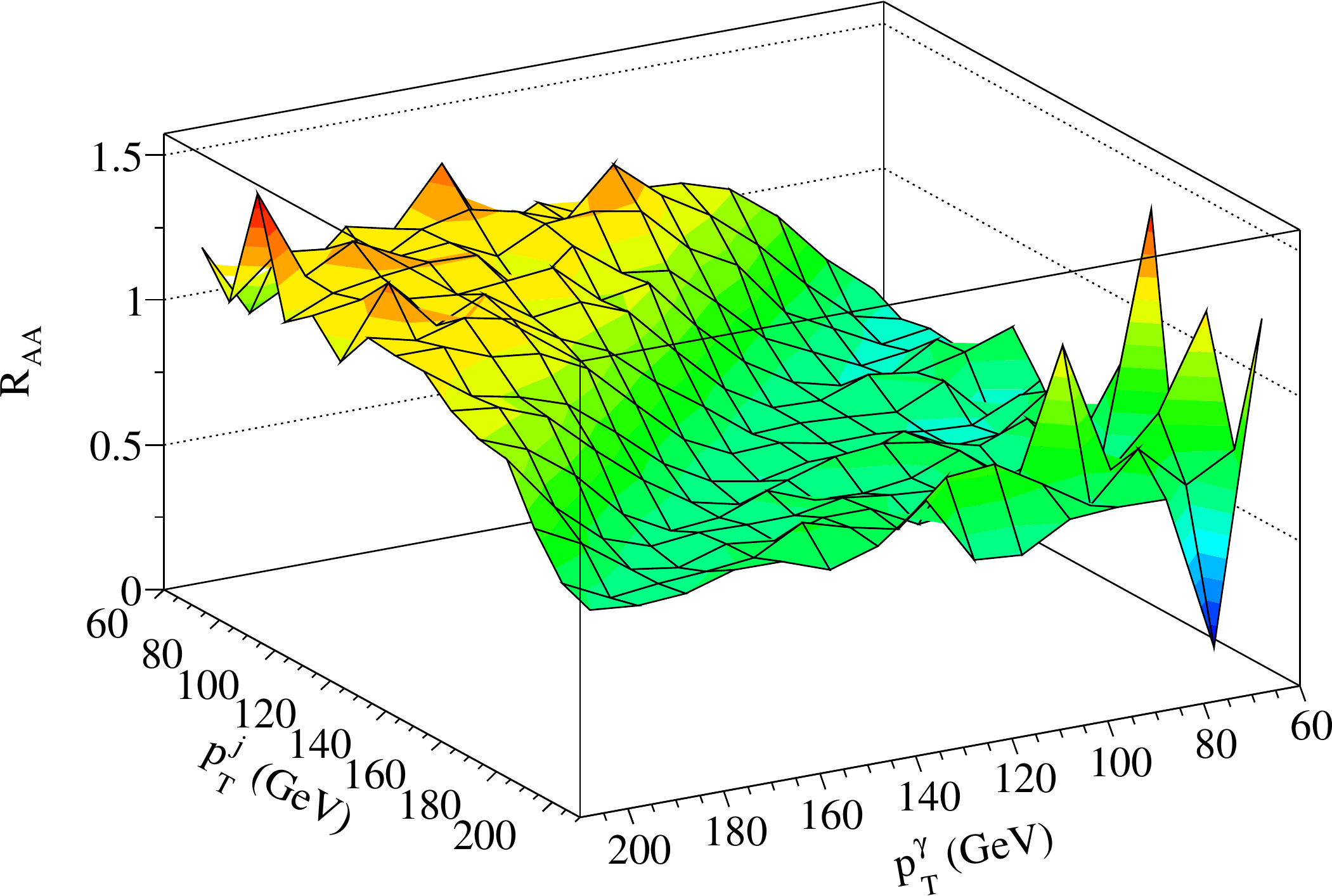, width=0.9\columnwidth}
\hskip 0.1 in 
\psfig{file=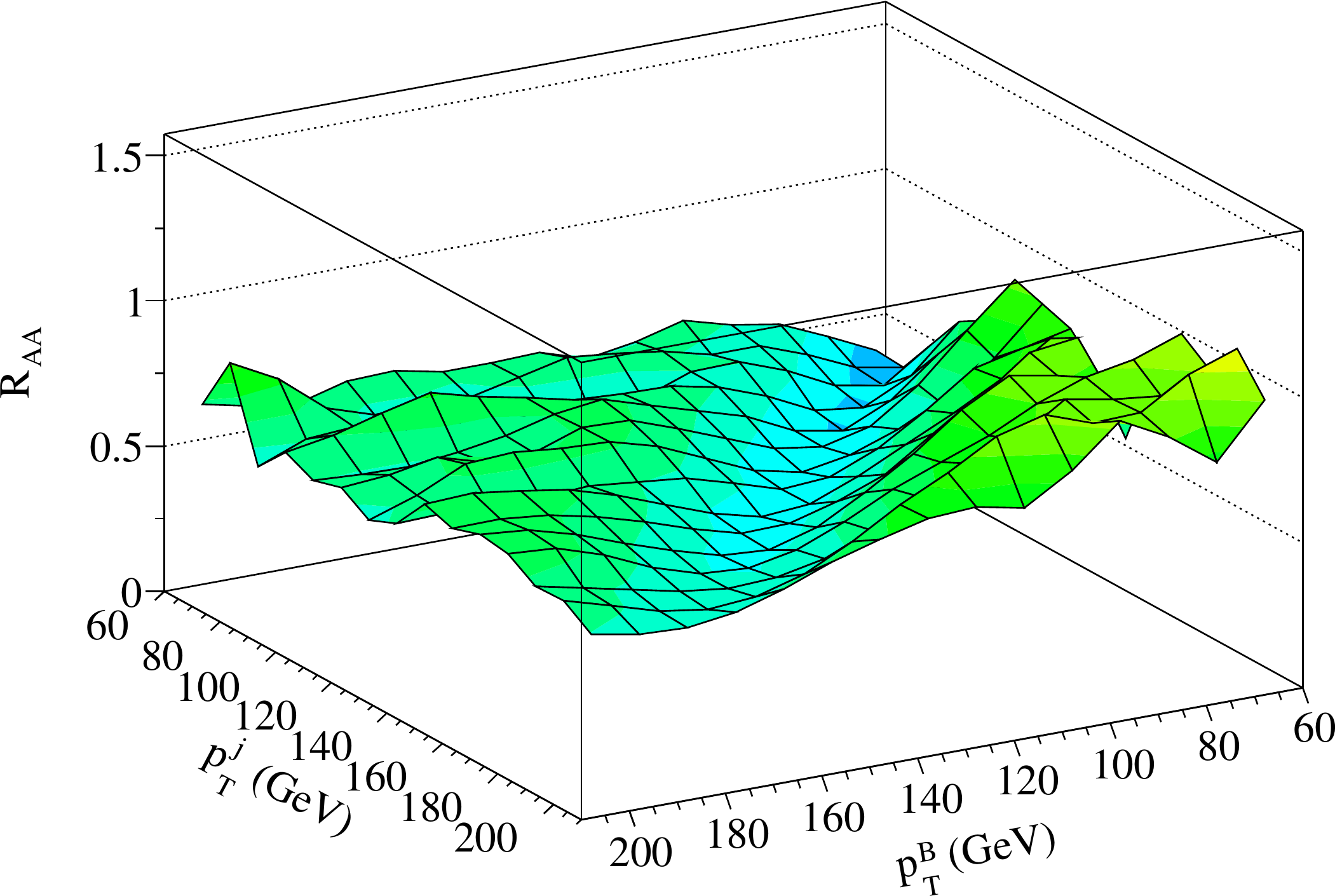, width=0.9\columnwidth}
\caption{The nuclear modification factor $R_{AA}$ for photon tagged (top) and B-meson-tagged (bottom) b-jets production in Pb+Pb collisions at LHC $\sqrt{s}=5.1$ TeV. }
\label{Raa-tagged-b}
\eef

Predictions for the nuclear modification factor $R_{AA}$ are presented in Fig. \ref{Raa-tagged-b} for photon-tagged b-jet (top) and B-meson-tagged b-jet (bottom), respectively. The largest suppression is observed in the diagonal region ($p_T^{\gamma, B} \approx p_T^j$) for both  tagged processes, and enhancement can exist in the region of $p_T^{\gamma} > p_T^j$ for photon-tagged b-jet, which is consistent with what was observed in photon-tagged light flavor jet production \cite{Dai:2012am}. As we have expected, due to the smaller contribution from gluon splitting mechanism in photon-tagged b-jets than inclusive b-jets, the suppression of photon-tagged b-jets is smaller than the quenching in inclusive b-jets when $p_T^{\gamma} < p_T^j$. However, in the B-meson-tagged b-jets case the overall suppression is stronger because both the parton that fragments into final state B-meson and b-jet lose energy, while the isolated photon escapes out of the medium without strong interaction. 

To further quantify the energy loss effect, we consider the tagged b-jet event asymmetry as a function of the asymmetry variable $z_{j\gamma/B}=p_T^j/p_T^{\gamma/B}$.  The momentum imbalance distributions are given in Fig. \ref{say}, for photon-tagged (top) and B-meson-tagged (bottom) b-jets, respectively.  For the photon-tagged b-jets, we see a moderate suppression of the cross section when $z_{j\gamma}\sim 1$, and the average asymmetry variable is shifted to smaller values. This arises from the energy loss of the b-jet side, while the isolated photon does not lose energy. For B-meson-tagged b-jets, stronger suppression is observed when $z_{jB}\sim 1$, due to both the partons on trigger and recoil side lose energy. Besides the suppression, in contrast to the photon-tagged case a slight increase of the asymmetry $z_{jB}$ is observed. This is due to the smaller energy loss on the recoiling b-jet side in comparison to the tagging B-meson side, which is different from the light dijet asymmetry enhancement ~\cite{He:2011pd}. 

\bef
\psfig{file=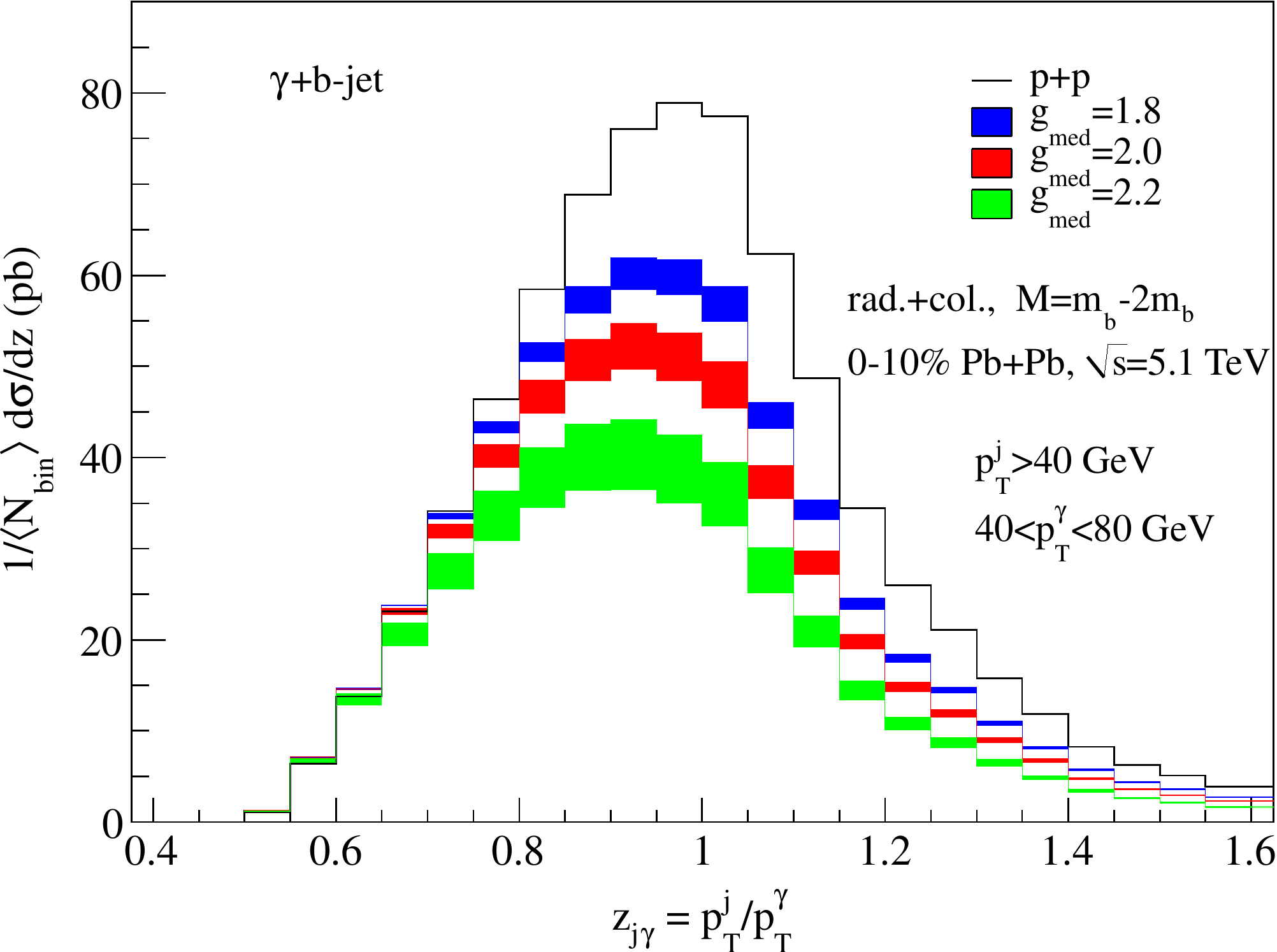, width=0.85\columnwidth}
\hskip 0.1 in 
\psfig{file=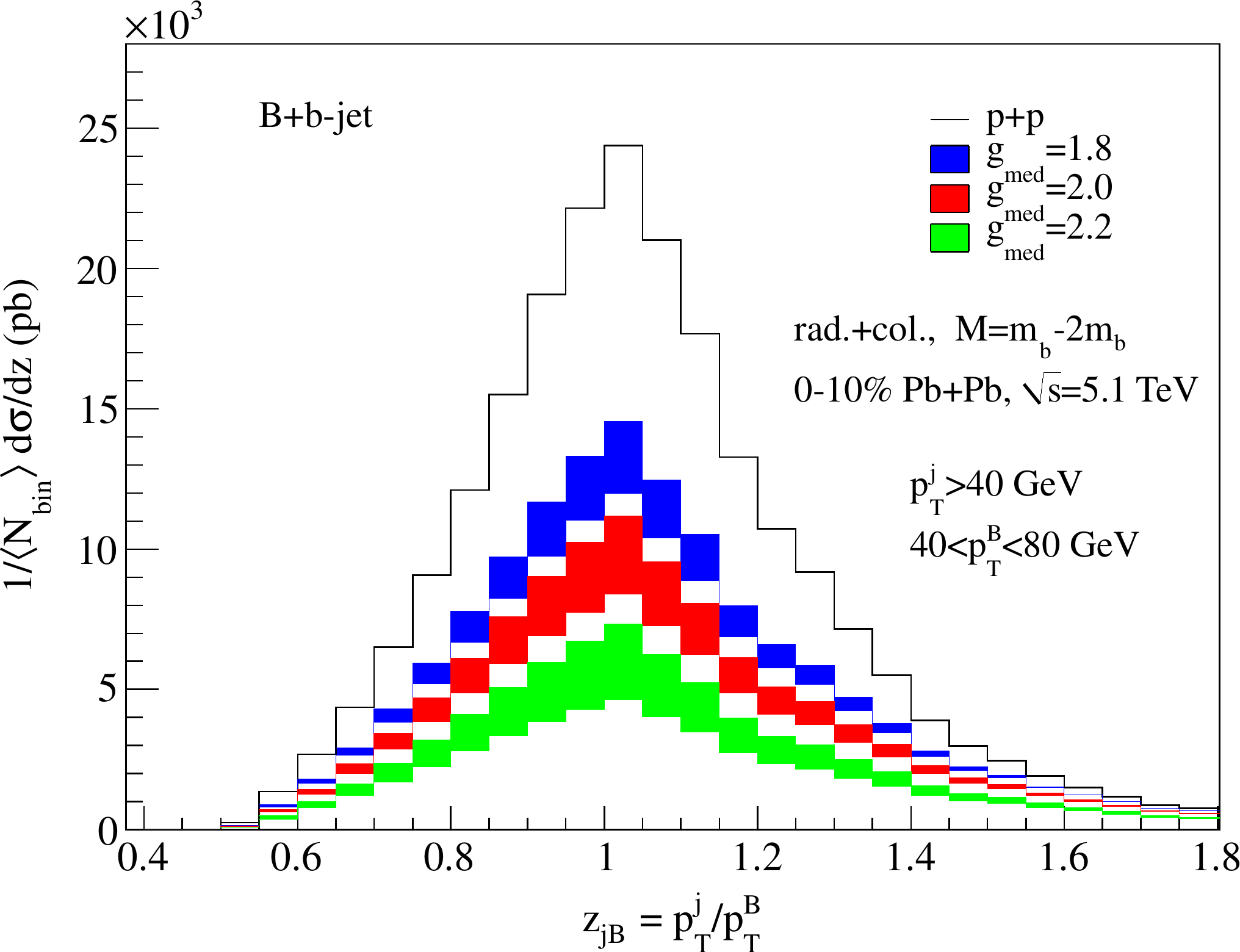, width=0.85\columnwidth}
\caption{The photon tagged (top) and B-meson tagged (bottom) b-jet asymmetry distribution for different coupling strength $g_{\rm med }= 1.8, 2.0, 2.2$. }
\label{say}
\eef

\section{Summary}
In summary, we presented theoretical predictions for the nuclear modification of the inclusive and tagged b-jet differential cross section in Pb-Pb collisions at the LHC. We use Pythia 8 simulation for the p+p baseline, and the GLV energy loss formalism to take into account QGP effects. The numerical results for inclusive b-jet production have been shown to agree well with with the subsequent CMS measurements. The magnitude of b-jet quenching is comparable to the observed attenuation of light jets in the large $p_T$ region. This can be understood form the disappearance of heavy quark mass effects in the high transverse momentum region, as well as the small fractional contribution of the b-jet that originates from the prompt b-quark.  

To better constrain the heavy-quark energy loss mechanism, we proposed to consider isolated-photon-tagged and B-meson-tagged b-jets production. In these processes the fractional contribution of  b-jets that originate from the prompt b-quark is greatly  increased. This is particularly pronounced in the B-meson-tagged case, where the b-quark fraction can reach up to $70\%$, which implies a direct connection to physical b-quark energy loss mechanism.  We presented predictions for both  the nuclear modification $R_{AA}$ and the related momentum imbalance distribution. Significant nuclear suppression is observed in the symmetric transverse momentum region ($p_T^j\sim p_T^{\gamma/B}$, or $z_{j\gamma /B}\sim 1$). Down-shift and up-shift of the asymmetry variable are observed for photon-tagged and B-meson tagged b-jet, respectively. By comparing such predictions to the upcoming experimental measurements at LHC, the tagged b-jet observables can provide us with unique new insights into heavy flavor dynamics in the hot dense medium. 

This research is supported by the US DOE Office of Science, and in part by the LDRD program at Los Alamos National Laboratory. 

\nocite{*}
\bibliographystyle{elsarticle-num}
\bibliography{martin}



\end{document}